\documentclass[12pt,preprint]{emulateapj}
\usepackage{graphicx}

\shorttitle{The Kepler-68 System}
\shortauthors{Stephen R. Kane}
\slugcomment{Submitted for publication in the Astrophysical Journal
  Letters}

\begin{document}

\title{Stability of Earth-mass Planets in the Kepler-68 System}

\author{Stephen R. Kane}
\affil{Department of Physics \& Astronomy, San Francisco State
  University, 1600 Holloway Avenue, San Francisco, CA 94132, USA}
\email{skane@sfsu.edu}

%%%%%%%%%%%%%%%%%%%%%%%%%%%%%%%%%%%%%%%%%%%%%%%%%%%%%%%%%%%%%%%%%%%%

\begin{abstract}

A key component of characterizing multi-planet exosystems is testing
the orbital stability based on the observed properties. Such
characterization not only tests the validity of how observations are
interpreted but can also place additional constraints upon the
properties of the detected planets. The {\it Kepler} mission has
identified hundreds of multi-planet systems but there are a few that
have additional non-transiting planets and also have well
characterized host stars. Kepler-68 is one such system for which we
are able to provide a detailed study of the orbital dynamics. We use
the stellar parameters to calculate the extent of the Habitable Zone
for this system, showing that the outer planet lies within that
region. We use N-body integrations to study the orbital stability of
the system, in particular placing an orbital inclination constraint on
the outer planet of $i > 5\degr$. Finally, we present the results of
an exhaustive stability simulation that investigates possible
locations of stable orbits for an Earth-mass planet. We show that
there are several islands of stability within the Habitable Zone that
could harbor such a planet, most particularly at the 2:3 mean motion
resonance with the outer planet.

\end{abstract}


%%%%%%%%%%%%%%%%%%%%%%%%%%%%%%%%%%%%%%%%%%%%%%%%%%%%%%%%%%%%%%%%%%%%

\section{Introduction}
\label{intro}

The characterization of exoplanetary systems and their host stars is
an ongoing effort, often requiring substantial telescope time. This is
particularly true of the exoplanet candidates from the {\it Kepler}
mission \citep{bor11a,bor11b,bat13,bur14,row15}, for which validation
of the candidates with relatively faint host stars and/or small
planets can be challenging. Fortunately, multiplicity in an exoplanet
system reveals that it has an exceptionally low probably of being a
false-alarm, allowing many of the multi-planet Kepler systems to be
validated \citep{lis14,row14}. Additional characterization of such
systems includes testing orbital stability for those in which the
planetary masses can be determined, either through measurable Transit
Timing Variations (TTVs) or radial velocity (RV) effects
\citep{li14,mah14}.

An additional aspect that can be tested for multi-planet systems is
the possible locations of dynamical stability for terrestrial planets
within the star's Habitable Zone (HZ). Examples of stability studies
include predicting planets in known exosystems
\citep{bar04,ray05,ray06} as well as studies of individual systems
such as HD~47186 \citep{kop09}, 70~Vir \citep{kan15}, and numerous
other systems \citep{men03,kop10}. These studies require a well
characterized host star that allows accurate HZ calculations to be
performed. One such confirmed Kepler system is that of Kepler-68,
published by \citet{gil13}. Kepler-68 is relatively bright, making it
suitable for RV follow-up and astroseismology studies. These have been
used to great effect resulting not only in well-determined fundamental
properties of the host star, but also the detection of a
non-transiting giant planet, bringing the total number of known
planets in the system to three. This introduces a fascinating system
that deserves further study into the dynamical limitations imposed by
the known orbits.

Here we present the results of an exhaustive stability analysis of the
Kepler-68 system along with HZ calculations and constraints on the
presence of an Earth-mass planet in the HZ. Section~\ref{system}
discusses the known stellar and planetary parameters of the system and
uses these to calculate the system HZ. Section~\ref{stab} describes
the orbital stability of the system and shows the results of varying
the inclination of the outer non-transiting
planet. Section~\ref{earth} combines the methodology of the previous
two sections by inserting a hypothetical Earth-mass planet at a large
range of semi-major axes and locating the islands of stability that
may exist with respect to the HZ. We provide concluding remarks and
discussion of future work in Section~\ref{conclusions}.

%%%%%%%%%%%%%%%%%%%%%%%%%%%%%%%%%%%%%%%%%%%%%%%%%%%%%%%%%%%%%%%%%%%%

\section{System Parameters and Habitable Zone}
\label{system}

The Kepler-68 (KOI-246; KIC 11295426) system was originally identified
in the first release of Kepler candidates \citep{bor11a,bor11b}. The
star was found to harbor two transiting planets, labeled 'b' and 'c',
with orbital periods of 5.399 and 9.605 days respectively. The
brightness of Kepler-68 ($K_p = 10$) enabled precise RV measurements
that revealed the presence of a third ('d') non-transiting planet in
an eccentric orbit ($e = 0.18$) with an orbital period of
$\sim$580~days \citep{gil13}. The combination of spectroscopic and
astroseismic modeling were used to determine stellar properties of
$T_\mathrm{eff} = 5793 \pm 74$~K, $M_\star = 1.079 \pm 0.051
\ M_\odot$, $R_\star = 1.243 \pm 0.019 \ R_\odot$, and $L_\star =
1.564 \pm 0.141 \ L_\odot$.  The resulting semi-major axes of planets
b, c, and d are thus 0.061, 0.091, and 1.4~AU respectively and their
masses are 8.3~$M_\oplus$, 4.8~$M_\oplus$, and 0.947~$M_J$
respectively. Full details of the stellar and planetary properties of
the system may be found in \citet{gil13}.

Using the stellar properties described above, we calculate the
boundaries of the HZ as described by \citet{kop13,kop14}. We further
use the terms ``conservative'' and ``optimistic'' (see \citet{kan13})
to label HZ regions that are based on assumptions regarding the amount
of time that Venus and Mars were able to retain liquid water on their
surfaces \citep{kop13}. The uncertainties in HZ boundaries were
demonstrated by \citet{kan14} to be largely impacted by the stellar
radius and subsequent luminosity uncertainties. Fortunately the
stellar parameters for Kepler-68 are known sufficiently well from
astroseismology measurements such that the HZ boundary uncertainties
are negligibly small. For further information on HZ calculations for
all known exoplanetary systems using the above described methodology,
we refer the reader to the Habitable Zone Gallery \citep{kan12}.

\begin{figure}
  \includegraphics[angle=270,width=8.5cm]{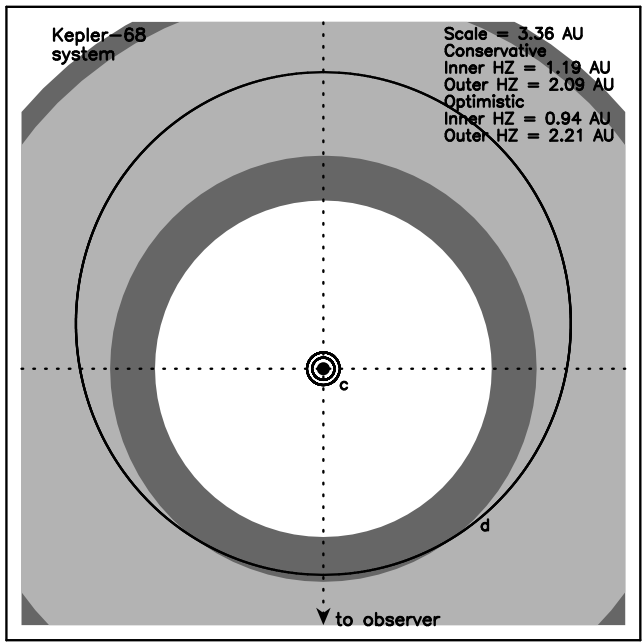}
  \caption{A top-down view of the Kepler-68 system showing the extent
    of the HZ and orbits of the planets calculated using the stellar
    and planetary parameters from \citet{gil13}. The physical scale
    depicted is 3.36~AU on a side. The conservative HZ is shown as
    light-gray and optimistic extension to the HZ is shown as
    dark-gray. The inner-most (unlabeled) orbit is that of planet b.}
  \label{hz}
\end{figure}

The calculated HZ boundaries for the Kepler-68 system are depicted in
Figure~\ref{hz}, along with the orbits of the three known planets.
The conservative HZ is shown as light-gray with inner and outer
boundaries located at 1.19 and 2.09~AU respectively. The optimistic
extension to the HZ is shown as dark-gray with inner and outer
boundaries of 0.94 and 2.21~AU respectively. A key aspect to notice in
Figure~\ref{hz} is that the measured orbit of the d planet spans a
wide range of star--planet separations throughout the HZ due to the
orbital eccentricity of the planet. It will therefore be significant
to determine if such an orbit excludes the presence of Earth-mass
planets in the HZ.

%%%%%%%%%%%%%%%%%%%%%%%%%%%%%%%%%%%%%%%%%%%%%%%%%%%%%%%%%%%%%%%%%%%%

\section{Orbital Stability and Inclination of Outer Planet}
\label{stab}

Considering the known planets in the Kepler-68 system include two very
close-in planets of moderate mass and a giant eccentric planet of
unknown orbital inclination, it is useful to test stability scenarios.
To do this, we calculated dynamical simulations using N-body
integrations with the Mercury Integrator Package, described in more
detail by \citet{cha99}. Our simulations made use of the hybrid
symplectic/Bulirsch-Stoer integrator and a Jacobi coordinate system,
since this coordinate system generally provides more accurate results
for multi-planet systems \citep{wis91,wis06} except in cases of close
encounters \citep{cha99}. All of the integrations were performed for a
simulation duration of $10^6$ years starting at the present epoch,
with results output in steps of 100 years. Occasional checks using
$10^7$~year simulations were used to verify that instability regions
beyond the $10^6$ timscale were not being missed. We used a time
resolution of 0.25~days to meet the recommended requirement of
choosing a timestep $1/20$ of the shortest orbital period
\citep{dun98}, $\sim$5.4~days in this case. A single simulation of the
system assuming that the system is approximately coplanar ($i =
90\degr$ for planet d) showed that the system is indeed stable over
long timescales for such a configuration.

\begin{figure*}
  \begin{center}
    \hspace{0.5cm}
    \includegraphics[angle=270,width=15.75cm]{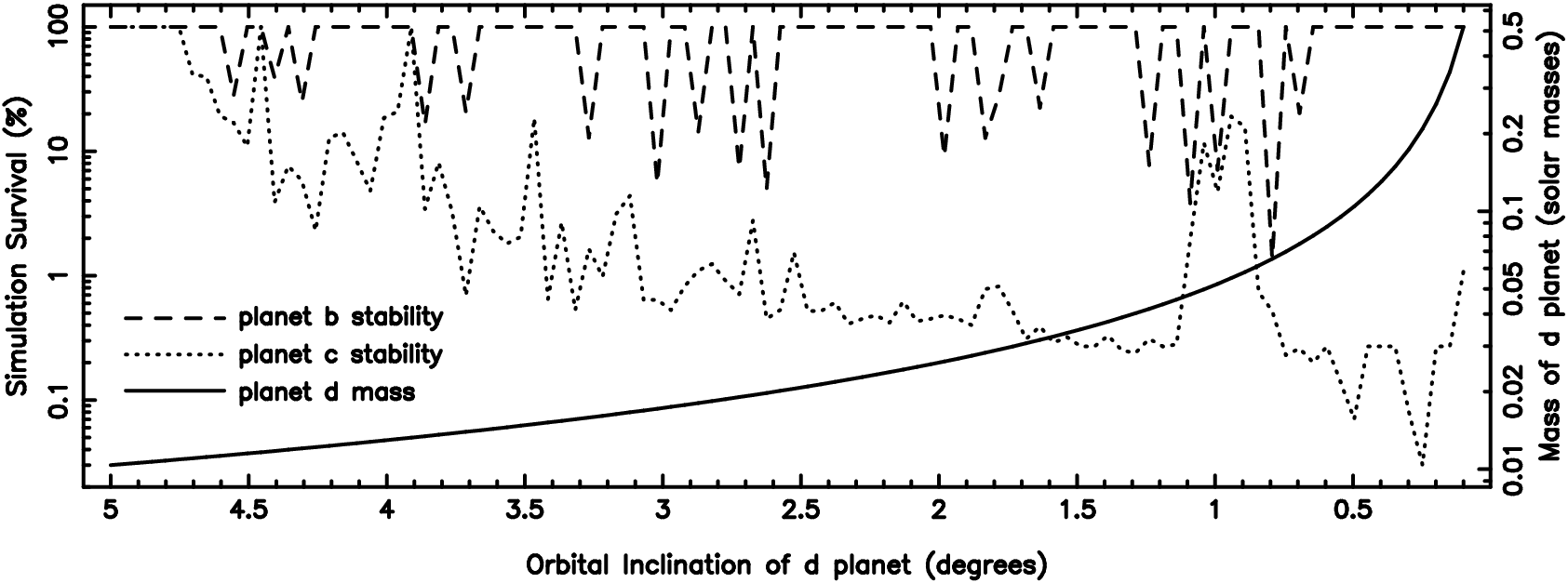} \\
    \includegraphics[angle=270,width=15.0cm]{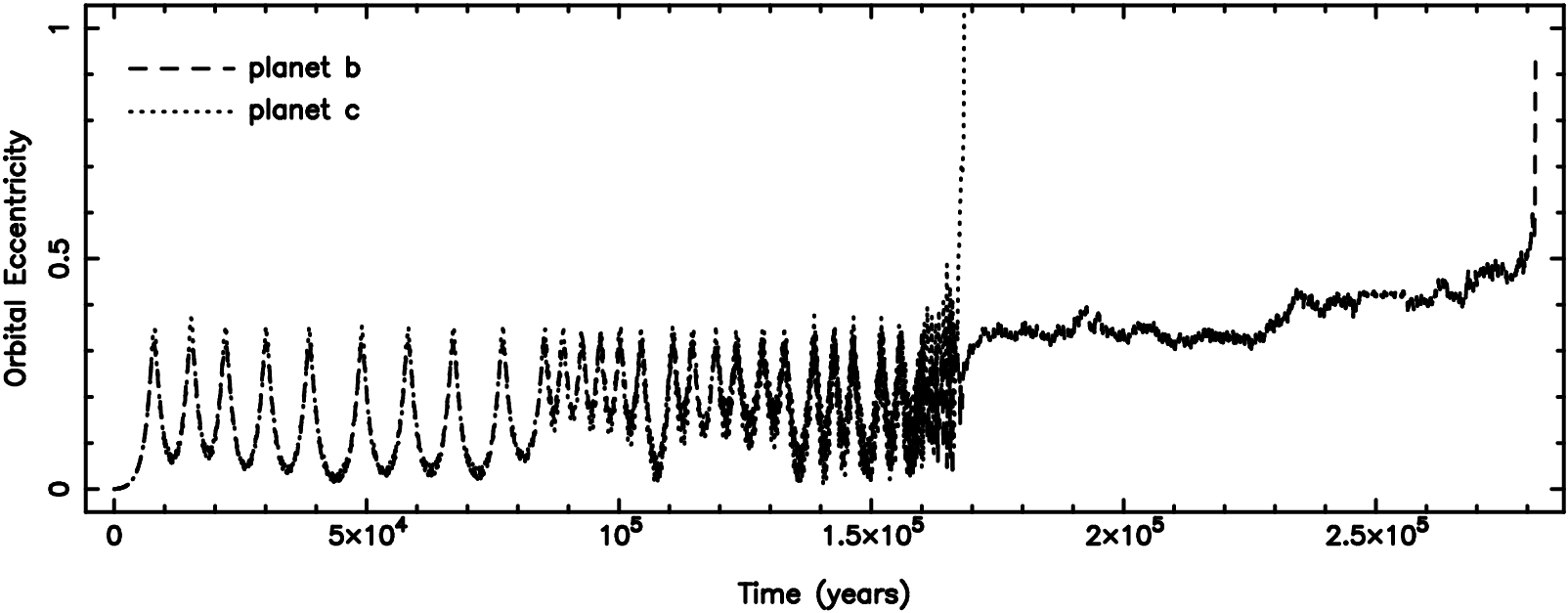}
  \end{center}
  \caption{Top: The percentage simulation survival of the two inner
    planets as a function of orbital inclination of planet d. The
    survival of planets b and c are shown as dashed and dotted lines
    respectively. Also shown is a solid line that represents the
    increasing true mass of planet d as the orbital inclination moves
    toward being face-on, with the mass values shown on the right-hand
    axis. Bottom: The orbital eccentricities of planets b and c as a
    function of time for a planet d orbital inclination of
    $4.55\degr$. The eccentricities of planets b and c are depicted as
    dashed and dotted lines respectively. The eccentricities of the
    two inner planets oscillate in identical ways until planet c is
    lost after $\sim$170,000~years. Planet b then survives in the
    system until $\sim$280,000~years.}
  \label{incecc}
\end{figure*}

To test the effect of planet d inclination, we performed a sequence of
stability simulations that vary the orbital inclination of the outer
planet from edge-on ($i = 90\degr$) to face-on ($i = 0\degr$) in steps
of $1\degr$. Since the true mass of the outer planet is the measured
mass divided by $\sin i$, the mass of the outer planet in the
simulations is adjusted accordingly. The stability simulations show
that the system remains stable throughout the full duration of the
simulation for all inclinations of planet d in the range $90\degr \leq
i \leq 5\degr$. For inclinations less than $5\degr$, the orbits of the
inner planets become significantly perturbed by planet d such that the
system stability is compromised. To increase the inclination
resolution in this region, we performed a further set of stability
simulations from $5\degr$ to $0\degr$ in steps of $0.05\degr$. The
survival of the inner planets as a function of planet d's orbital
inclination is shown in the top panel of Figure~\ref{incecc}. The
system is clearly catastrophically unstable within this range, where
the seemingly chaotic survival times are due to the sensitivity of the
simulations to initial starting conditions.

The range of true masses for planet d in the inclination range
$5\degr$--$0\degr$ is 0.01--0.5 solar masses, shown as a solid line in
Figure~\ref{incecc}. Inclinations less than $1\degr$ move the mass of
the d component into the stellar regime. In addition to the
instability reasons, it is highly unlikely that the d component is
stellar in nature due to the expected evidence from follow-up
observations. The {\tt BLENDER} analysis developed by \citet{tor11}
and used by \citet{gil13} did not indicate a centroid shift that is
characteristic of a false-positive, nor did the acquired spectra show
signs of lines shifts due to a stellar companion.

The bottom panel of Figure~\ref{incecc} shows the change in
eccentricities of planets b and c as a function of time for the case
where planet d's inclination is $4.55\degr$. This is shown as an
example of the orbital degradation that occurs at the threshold of
instability. The inner planets oscillate with the same frequency until
planet c is removed from the system at a time of $\sim$170,000~years,
after which planet b survives until $\sim$280,000~years. To be clear,
the removal of planets in this context does not mean that they are
ejected from the system. The inner planets are unlikely to be able to
escape the gravitational potential well of the host star and so the
planets will be consumed by the star rather than ejected. Since the
radius of star is 0.0058~AU, the eccentricities for which this will
occur for planets b and c are 0.90 and 0.94 respectively.

We have shown that the inclination of the outer planet has to be
unrealistically small in order to render the system unstable. Indeed
most of the Kepler systems have been found to be consistent with the
prevalence of coplanarity amongst multi-planet systems
\citep{fan12}. A consequence of this is that there are large stability
regions that exist within the system as possible locations for further
planets, as investigated in the next section.

%%%%%%%%%%%%%%%%%%%%%%%%%%%%%%%%%%%%%%%%%%%%%%%%%%%%%%%%%%%%%%%%%%%%

\section{Constraints on the Presence of an Earth-mass Planet}
\label{earth}

As noted earlier, the stellar properties of the Kepler-68 host star
are relatively well defined. This presents an opportunity to examine
the possibility that other terrestrial planets may exist in the system
beneath the detection limits of current data. To test the viability of
such a scenario, we conducted an exhaustive set of simulations that
place an Earth-mass planet with a circular orbit at various locations
within the system. We assumed that the planetary orbits are
approximately coplanar including the known outermost planet d.

\begin{figure*}
  \begin{center}
    \includegraphics[angle=270,width=15.5cm]{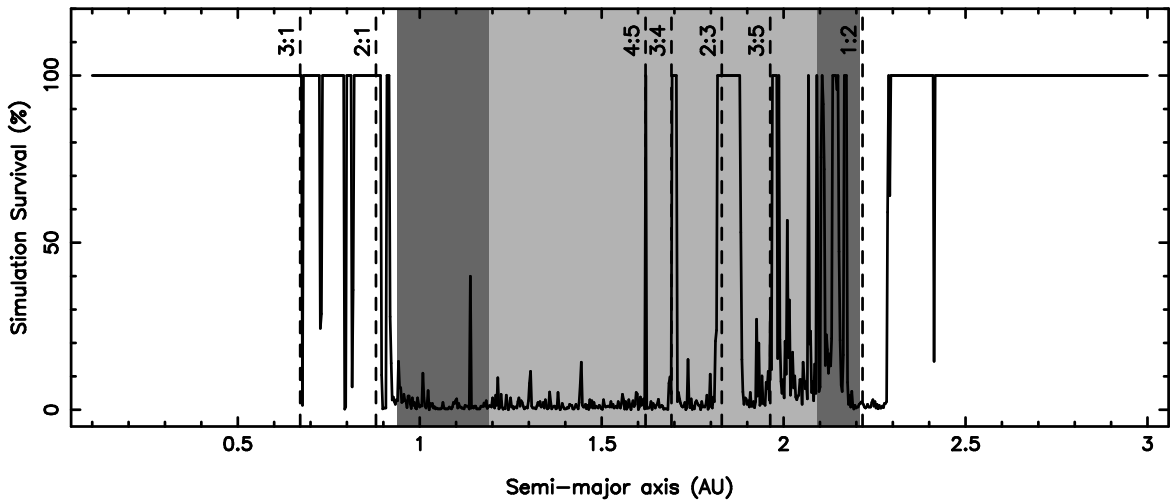}
  \end{center}
  \caption{The orbital stability of a hypothetical Earth-mass planet
    as a function of semi-major axis in the Kepler-68 system. The
    orbital stability on the vertical axis is expressed as the
    percentage simulation survival for each semi-major axis where the
    position of the planet was tested. The system is assumed to be
    coplanar and the orbit of the Earth-mass is assumed to be
    circular. The light-gray and dark-gray regions represent the
    conservative and optimistic HZ regions, as per
    Figure~\ref{hz}. The vertical dashed lines indicate the locations
    of the mean-motion orbital resonances with planet d.}
  \label{earthfig}
\end{figure*}

Using the system configuration described above, we performed 1,000
simulations that increment the semi-major axis of the simulated planet
between 0.1 and 3.0~AU. The inner boundary of 0.1~AU was chosen to lie
just beyond the location of planet c. Each simulation had a duration
of $10^6$ years, as described in Section~\ref{stab}. Shown in
Figure~\ref{earthfig} are the results of the simulation, where the
percentage of the simulation that the inserted planet remained in the
system is plotted against the semi-major axis of the planet. The
light-gray and dark-gray regions represent the conservative and
optimistic HZ regions respectively.

Given that planet d has an eccentricity of $e = 0.18$ and a semi-major
axis $a = 1.4$, it is not surprising that there are vast swathes of
instability that lie throughout the HZ. The periastron and apastron
distances of planet d are 1.15 and 1.65~AU respectively. The mass of
the planet yields a Hill radius of $\sim$0.1~AU, consistent with the
boundaries of the instability regions seen in Figure~\ref{earthfig}
\citep{mar82,gla93}. The vertical dashed lines show the locations of
the mean motion orbital resonances (MMR) with planet d. As described
by \citet{ray08}, some MMR locations are preferred to others and
indeed there are MMR locations that are destabilizing regions (such as
the 1:2 MMR) rather than locations for MMR trapping. Although it seems
that the presence of planet d excludes the possibility of a
terrestrial planet in the Kepler-68 HZ, there lies an extended region
of stability within 1.8--1.9~AU that corresponds to the 2:3 MMR
location. This island of stability lies well within the conservative
HZ for the system. For an Earth-mass planet, the RV amplitude within
the stability island would be $\sim$6~cm\,s$^{-1}$, beyond the reach
of current instruments for a star so faint. Furthermore, an
Earth-radius planet would have a transit depth of $\sim$55~ppm with an
orbital period of $\sim$880~days at 1.85~AU. Such a signature would
occur at most twice within the {\it Kepler} time series, presenting
significant detectability challenges.

%%%%%%%%%%%%%%%%%%%%%%%%%%%%%%%%%%%%%%%%%%%%%%%%%%%%%%%%%%%%%%%%%%%%

\section{Conclusions}
\label{conclusions}

The necessity of follow-up for Kepler candidates is clearly vital for
both confirming their planetary status as well as understanding the
host stars. Kepler-68 is one of the few cases where non-transiting
planets have been discovered through such follow-up activities,
revealing an intriguing system with an eccentric giant planet (planet
d). We have calculated the HZ for this system and have shown that the
d planet carves a path right through the HZ.

When non-transiting planets are found, it lays open the subject of how
the system stability depends on the inclination of such planets. To
resolve this issue, we have conducted extensive N-body stability
simulations that show the system is stable for all inclinations of the
d planet from $90\degr$ to $\sim$$5\degr$. The d planet begins to
adopt stellar properties for inclinations less that $1\degr$ which
reinforces the exclusion of such an inclinations since the companions
would undoubtedly have been detected in the follow-up data described
by \citet{gil13}.

We further investigated the locations within the system where a
terrestrial planet of one Earth-mass could potentially maintain a
stable orbit. We showed how planet d is a particularly destabilizing
factor within the system HZ, apart from an island of stability between
1.8--1.9~AU and various other narrow locations where MMR trapping can
occur. There are many locations outside of the HZ where such a
hypothetical planet could be harbored without risk of gravitational
encounters that would render its continued orbit untenable. The other
reason that the presence of the giant planet is not necessarily a
deterrent to habitability is because such a planet could harbor
terrestrial moons that may themselves be habitable
\citep{for13,hin13,hel14}. However, the frequency of giant planets
within the HZ is clearly a topic that cannot be ignored in subsequent
searches for terrestrial HZ planets.

%%%%%%%%%%%%%%%%%%%%%%%%%%%%%%%%%%%%%%%%%%%%%%%%%%%%%%%%%%%%%%%%%%%%

\section*{Acknowledgements}

The author would like to thank Gregory Laughlin and Sean Raymond for
useful discussions on the stability simulations. Thanks are also due
to Ronald Gilliland and the anonymous referee, whose feedback improved
the quality of the paper. This research has made use of the following
archives: the Exoplanet Orbit Database and the Exoplanet Data Explorer
at exoplanets.org, the Habitable Zone Gallery at hzgallery.org, and
the NASA Exoplanet Archive, which is operated by the California
Institute of Technology, under contract with the National Aeronautics
and Space Administration under the Exoplanet Exploration Program. The
results reported herein benefited from collaborations and/or
information exchange within NASA's Nexus for Exoplanet System Science
(NExSS) research coordination network sponsored by NASA's Science
Mission Directorate.

%%%%%%%%%%%%%%%%%%%%%%%%%%%%%%%%%%%%%%%%%%%%%%%%%%%%%%%%%%%%%%%%%%%%

\end{document}